\documentclass{article}

\usepackage{amsmath}

\usepackage{lipsum}
\usepackage{amsthm}
\usepackage{float}
\usepackage{amsthm}
\theoremstyle{definition}
\usepackage{subcaption}
\newtheorem{example}{Example}
\usepackage{booktabs} 
\usepackage{listings}
\usepackage{array} 
\usepackage{multirow} 

\usepackage{arxiv}

\usepackage[utf8]{inputenc} 
\usepackage[T1]{fontenc}    
\usepackage{hyperref}       
\usepackage{url}            
\usepackage{booktabs}       
\usepackage{amsfonts}       
\usepackage{nicefrac}       
\usepackage{microtype}      
\usepackage{lipsum}
\usepackage{graphicx}
\graphicspath{ {./images/} }

\title{Understanding User Preference - Comparison between Linear and Directional Top-K Query results}

\author{
 Xiaolei Jiang \\
  School of Informatics\\
  Università della Svizzera italiana\\
  6900 Lugano, Switzerland \\
  \texttt{xiaolei.jiang@usi.ch} \\
  \\
}

\begin{document}
\maketitle
\begin{abstract}
This paper investigates user preferences for \textit{Linear Top-k Queries} and \textit{Directional Top-k Queries}, two methods for ranking results in multidimensional datasets. While \textit{Linear Queries} prioritize weighted sums of attributes, \textit{Directional Queries} aim to deliver more balanced results by incorporating the spatial relationship between data points and a user-defined preference line. The study explores how preferences for these methods vary across different contexts by focusing on two real-world topics: used cars (e-commerce domain) and football players (personal interest domain).

A user survey involving 106 participants was conducted to evaluate preferences, with results visualized as scatter plots for comparison. The findings reveal a significant preference for directional queries in the used cars topic, where balanced results align better with user goals. In contrast, preferences in the football players topic were more evenly distributed, influenced by user expertise and familiarity with the domain. Additionally, the study demonstrates that the two specific topics selected for this research exhibit significant differences in their impact on user preferences.

This research reveals authentic user preferences, highlighting the practical utility of Directional Queries for lifestyle-related applications and the subjective nature of preferences in specialized domains. These insights contribute to advancing personalized database technologies, guiding the development of more user-centric ranking systems.
\end{abstract}


\section{Introduction}
In today’s digital landscape, numerous applications help users easily access tailored results across diverse areas like e-commerce, scientific databases, web searches, and multimedia systems \cite{ppt}. For instance, in e-commerce, when users type in keywords or phrases, the search engine applies specific criteria—such as price, delivery cost, and sales volume—to rank and display the “best” k results, where "k" represents the preferred number of items to view. The criteria used (like price, delivery cost, and volume) are the key attributes that shape the ranking of each item.

The most common technique behind supporting these rankings is the linear top-k query\cite{DBLP:journals/csur/IlyasBS08}, where a scoring function that aggregates multiple attribute values with predefined weights given by users. And then base on the scores, rank the highest or lowest k numbers results as output.

Despite its effectiveness, linear top-k querying has limitations. A key drawback is its tendency to overemphasize extreme values in one attribute, which can lead to irrelevant or unbalanced results that may not align with user preferences. Additionally, setting appropriate weights for multi-dimensional criteria is a complex task. Small changes in weights can lead to significant differences in ranking outcomes, impacting the relevance of results.\cite{Flexible_Skylines,2021marrying,Regret-Minimizing,Soliman2011Ranking}

As the limitations of traditional linear top-k querying become more evident, a promising new method called the Directional Query has emerged to more align with user preferences. This method introduces a transformative concept to the scoring function: it not only calculates a weighted sum of each object’s attributes but also considers the spatial relationship between the object’s points and a preference line specified by the user. By measuring the distance of each result from this preference line, directional queries allow the query to prioritize results that are closer to the desired direction.

Although directional queries are recognized as an innovative method for ranking query results, but given the extensive applications of top-k queries, it is essential to consider how users respond to and evaluate this new directional query method. It raises an essential question: how do users perceive and respond to this advanced query method? Does the directional query effectively deliver results that align with users' interests and expectations? This master’s paper proposes to investigate this question by surveying real users to understand their perspectives on the difference between Linear and Directional Query results. By using a structured questionnaire, this study will capture user perspectives on the two methods across different contexts.Ultimately, the findings will reveal where directional queries best align with user preferences, offering critical insights for enhancing database query technologies.

And here is the summary of this project covered in this paper: 

\begin{itemize}
    \item Section 1 will introduce the concepts of Linear Top-k Queries and Directional Top-k Queries, providing a theoretical foundation and outlining the research objectives and motivation behind this study.
\end{itemize}

\begin{itemize}
    \item Section 2 will review recent and related works on top-k queries, user preference studies, and advancements in query ranking methods, offering a context for the research.
\end{itemize}

\begin{itemize}
    \item Section 3 will describe the methodology used in this research, including the design of the structured questionnaire, the datasets selected for analysis, and the approach for data collection and preprocessing.
\end{itemize}

\begin{itemize}
    \item Section 4 will present and discuss the results of the user preference survey, analyzing the findings in terms of topic type, user demographics, and knowledge levels, and interpreting their significance in relation to the research hypotheses
\end{itemize}

\begin{itemize}
    \item Section 5 will discuss the limitations of this research, present future directions such as expanding the sample size, increasing topic diversity, and incorporating more realistic query scenarios, and conclude the study by summarizing the key findings and the implications.
\end{itemize}

\subsection{Preliminary concepts}

Before delving into the design of questionnaires and detailed methodological approaches, it is essential to provide an overview of foundational concepts related to Linear and Directional Queries. This section will offer a brief introduction to these concepts to establish a theoretical foundation that will inform and contextualize the discussions in the following sections.

To begin with, the first concept need to introduce is the relational schema \( R(A_1, \dots, A_d) \), where each \( A_i \) represents an attribute and \( d \geq 1 \). All attribute values will be numerical and, without loss of generality, normalized in the interval \( [0, 1] \). From now on, the reference dataset (collection) will be identified with an instance \( r \) over \( R \). The tuples \( t \in r \) will therefore be equivalent to \( d \)-dimensional vectors. As such, these tuples can also be seen as points in a \( d \)-dimensional space, which will be referred to as the attribute space. Accordingly, the formulation \( t_i \) will be used as shorthand for \( t[A_i] \) throughout this work. Finally, the attribute values will be considered better the smaller they are.

The second essential concept is the top-k query, which is central to this paper. A top-k query employs a scoring function \(S\) that calculates a score for each object based on specific criteria and attributes.
The most commonly used family of functions for selecting scoring functions is the \( L_p \) family of weighted norms:

\vspace{1em}
\textbf{Definition 1.1. \( L_p \) Norms}

\[
L^W_p(t) = \left( \sum_{i=1}^d w_i t_i^p \right)^{\frac{1}{p}}, \quad p \in \mathbb{N} \tag{1.1}, \quad \forall i \; 0 \leq w_i \leq 1
\]

In this definition, \( W = (w_1, \dots, w_d) \) is a normalized weight vector, and the condition is \( \sum_i w_i = 1 \). This weight vector represents the importance assigned to each attribute, embodying one of the key aspects of top-k queries: personalization. 
By ranking these tuples in descending or ascending order, the query returns the k highest (or lowest) scoring objects. Here, $k$ is a user-defined parameter that determines the number of results returned, making top-k queries highly adaptable across various domains. 

In the following two subsections, we will provide a detailed explanation of linear top-k queries and directional top-k queries, including their definitions and the key differences between them.

\subsubsection{Linear Top-k query}

Linear top-k queries, also known as the \( L_1 \) norm, use a scoring function where \( p = 1 \) in Definition 1.1. This approach is widely used to retrieve relevant results from multidimensional databases. It effectively supports user preferences by allowing customized attribute weights within the scoring function. Additionally, linear top-k queries give users control over the output size by specifying the value of \( k \), which determines the number of results to be returned. The following example illustrates the linear top-k scoring function in action.

\begin{example}
Penny, a recent graduate, has received a job offer in another city and needs to find an apartment nearby so she can commute within a reasonable time. Initially unfamiliar with the local rental market, she considers two attributes---price (\( P \)) and distance (\( D \)) from her workplace---as equally important, each with a weight of 50\%. This gives each apartment a score \( S = p \times 0.5 + d \times 0.5 \), where \(\ p \) is the normalization of \(\ P \) and \( d \) is the normalization of \( D \). However, upon browsing listings, Penny realizes that rental prices are significantly higher than she anticipated due to the city’s international appeal. To find a more affordable place, she decides to adjust her priorities by giving price a weight of 70\% and distance 30\%, resulting in a new scoring function \( S = p\times 0.7 + d \times 0.3 \). As both price and distance are attributes that benefit from being lower, she sorts the apartments in ascending order based on their scores and selects her top three or five options.

The example above highlights the importance of the weight vector, as it influences each object's score and ultimately affects the final ranking results. The Figure 1 demonstrate the impact of different weight vector combinations.

\begin{figure}
    \centering
    \includegraphics[width=1\linewidth]{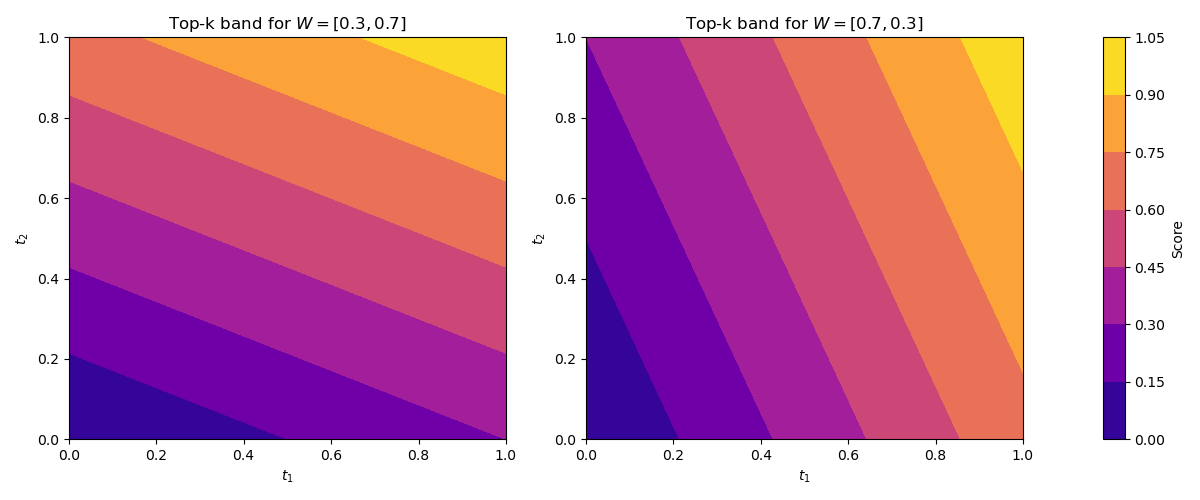}
    \caption{Top-K band on different weight vectors}
    \label{fig:enter-label}
\end{figure}

\end{example}

The linear top-k scoring function of \(L\) norm is when \(p\) =1 :

\vspace{1em}
\textbf{Definition 1.2.}

\[
L_1 = \left\{ f \ \middle|\ f(t) = \sum_{i=1}^d w_i t[i] \right\}\tag{1.2}
\]

where for a weight vector \(\mathbf{w} = \langle w_1, \dots, w_d \rangle\), 
\( \sum_{i=1}^d w_i = 1 \land \forall i, \ w_i \in [0, 1] \)

After scoring each object, they are ranked in ascending or descending order to retrieve the desired \( k \) results. In this paper, an ascending order is applied, which returns tuples that minimize the scoring function. Here, the origin \((0, \dots, 0)\) is treated as the optimal point, making it the reference for evaluating relevance. This sorting-based approach for top-k queries is not only simple but also computationally efficient, with a complexity of \(O(n \log k)\) for a single, unordered relation. Moreover, when joining multiple relations, established algorithms like TA and NRA can be employed, adding only a minor sub-linear overhead to the process.\cite{fagin2002optimal}

In practice, the need for such queries arises in various modern contexts, including e-commerce, scientific databases, web search, and multimedia systems \cite{ppt}. However, linear top-k queries also present some limitations that may not fully meet user expectations due to the characteristics of linear scoring functions. When dealing with a high number of attributes, users often find it challenging to accurately define the subtle weight differences for each attribute. Additionally, linear top-k query results tend to focus on points along the convex hull of the dataset—representing the smallest convex polygon or polyhedron that encompasses all data points—without reaching into the interior. This characteristic means that even with large \(k\) values intended to retrieve more relevant results, certain interior points that may be important could be missed \cite{onion}. At the same time, some balanced results may be overlooked, resulting in an unbalanced output.\cite{DBLP:journals/paccmod/CiacciaM24}

\subsubsection{Directional Top-K query}

The concept of directional top-k queries, proposed in recent years by \cite{DBLP:journals/paccmod/CiacciaM24} and \cite{thesis_dir}, offers a novel enhancement to traditional linear top-k queries, aiming to deliver results that better align with user preferences. To achieve more balanced outcomes for users without increasing the complexity of query configurations, the directional query retains the structure of the linear top-k scoring function. Yet, it incorporates the concept of \textit{distance} to the user-defined \textit{preference line} (PL). This addition means that, besides calculating the weighted sum of the attribute values, the query also considers each point’s distance to the PL. In the simplest case, where each attribute is assigned equal weight, the PL aligns with the diagonal, and objects closer to this line represent those most closely aligned with the user's preferences.\cite{DBLP:journals/paccmod/CiacciaM24}.
Continuing with \textit{Example 1} of Penny, which illustrates the limits of linear top-k query and the needs for directional query:

\begin{example}
    Penny has now been working in this city for three years and has recently been promoted to a senior position in her team. With her higher salary, she can afford a more expensive apartment and no longer wants to endure a lengthy commute. She decides to search for a new place and considers both price (P) and distance (D) equally important. However, the limitations of linear top-k queries become apparent---when Penny searches for the top 10 options, the results often include properties with extreme attribute values, such as apartments extremely close to her workplace but with exorbitant rents, or options with reasonable rents but far from her office. This inefficiency frustrates her, as expanding the search scope requires considerable time and effort to find a truly balanced result.
\end{example}

Due to the limitations of linear top-k, this query method no longer meets Penny’s needs, as she requires a more balanced result between the two attributes. By incorporating distance calculations, directional queries can help her achieve the balanced outcomes she seeks.

In directional query, to compute the distance, first define the preference line(PL): 

\vspace{1em}

\textbf{Definition 1.3.} The \textit{preference line} \( \text{PL}(\mathbf{w}) \) associated with the weight vector \( \mathbf{w} = \langle w_1, \dots, w_d \rangle \) is the set:
\[
\text{PL}(\mathbf{w}) = \left\{ \langle \bar{w}_1 x, \dots, \bar{w}_d x \rangle \ \middle|\ x \geq 0 \right\},\tag{1.3}
\]
where \( \bar{w}_i = \frac{1}{w_i} \) for \( 1 \leq i \leq d \).

To compute the distance between data point \( t \) and the preference line, it applies the Euclidean distance formula, \text{Dist}(t, \text{PL}(w)). This involves calculating, for each attribute, the difference between the data point's value \( t[i] \) in that dimension and its projected value along the preference line. This difference quantifies how far the data point deviates from the user’s ideal balance in each dimension, represented by:
\[
t[i] - \bar{w}_i \frac{\sum_{j=1}^d \bar{w}_j t[j]}{\sum_{j=1}^d \bar{w}_j^2}
\]

Then square each of these differences, sum them across all dimensions, and take the square root of the total to derive the Euclidean distance in Definition 1.4. A smaller distance implies that the data point better aligns with the user's preferences.

\vspace{1em}
\textbf{Definition 1.4.} The distance between a data point \( t \) and the preference line \( \text{PL}(w) \) is defined as:
\[
\text{Dist}(t, \text{PL}(w)) = \sqrt{\sum_{i=1}^d \left( t[i] - \bar{w}_i \frac{\sum_{j=1}^d \bar{w}_j t[j]}{\sum_{j=1}^d \bar{w}_j^2} \right)^2}\tag{1.4}
\]

The two components of the directional query, as outlined in Definition 1.5, are now established. This query method includes two key elements: the first is the weighted sum of attribute values based on a weight vector, and the second is the distance to the preference line defined by \( w \).

The family \( \text{DIR} \) of scoring functions for directional queries is defined as follows:

\vspace{1em}
\textbf{Definition 1.5.} The family \( \text{DIR} \) of scoring functions for directional queries is defined as:
\[
\text{DIR} = \left\{ f \ \middle| \ f(t) = \beta \sum_{i=1}^d w_i t[i] + (1 - \beta) \ \text{Dist}(t, \text{PL}(w)) \right\},\tag{1.5}
\]

Here, \( \beta \in [0, 1] \) acts as a tuning parameter that controls the trade-off between the weighted sum and the distance components. Modifying \( \beta \) allows the query to transition between two distinct modes: one where it behaves as a fully linear top-\( k \) query and another where it is driven entirely by distance. In the experiments in \cite{DBLP:journals/paccmod/CiacciaM24}, by lowering \( \beta \), the ranking prioritizes tuples near the preference line, enhancing the ranking of tuples with more balanced attribute values. And based on the tests done by \cite{thesis_dir}, the best parameter value is \( \beta \) = 0.66 in the majority of the cases. 

\begin{figure}[H] 
    \centering
    \includegraphics[width=0.5\linewidth]{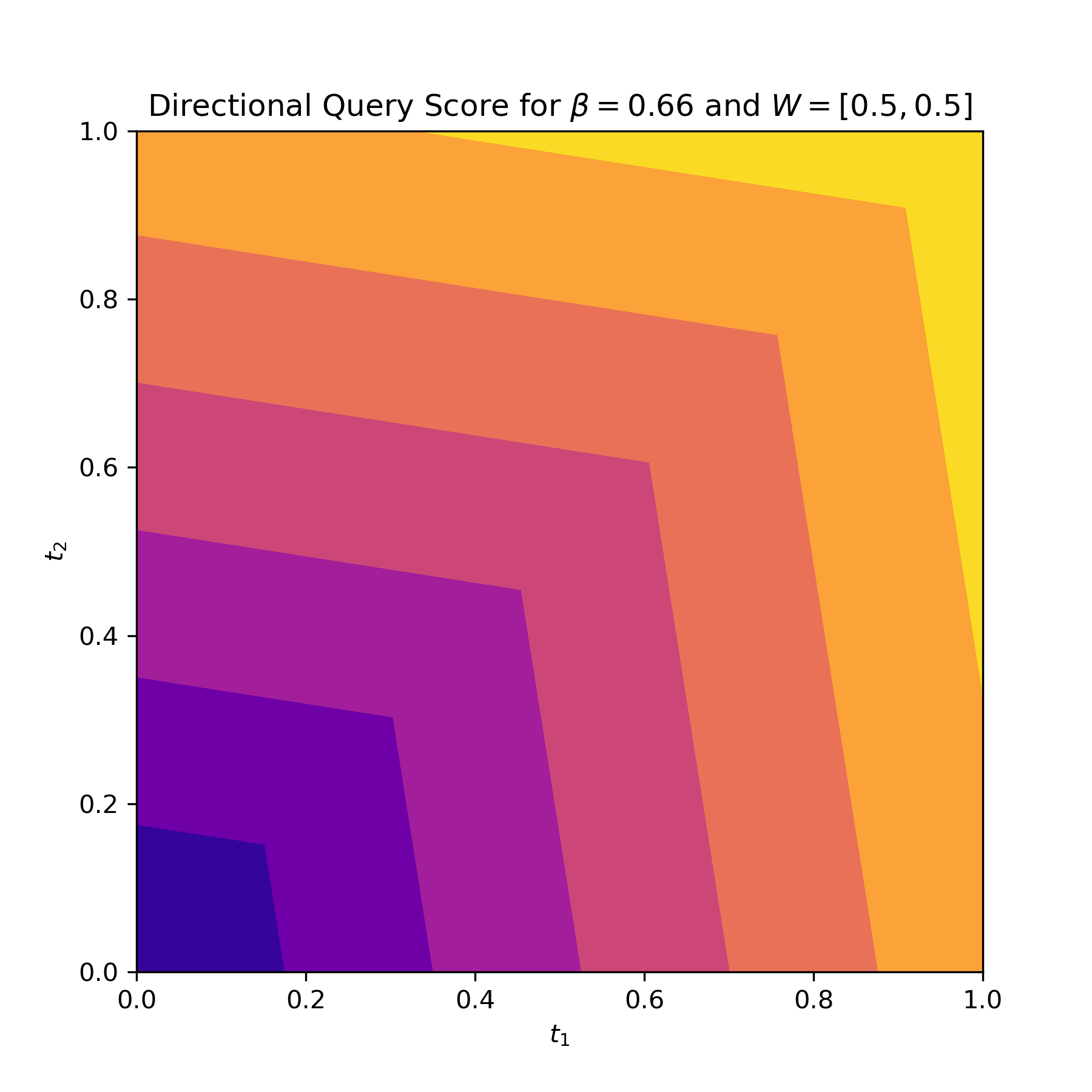}
    \caption{Directional bank when \(\beta\)=0.66 and W =[0.5,0.5]}
    \label{fig:enter-label}
\end{figure}

\section{Related Work}
The core of this paper is to understand user preferences regarding query results, with a focus on the linear top-k query and directional query approaches. Since directional queries were introduced only recently, there has been minimal research on how users respond to this new approach. A preliminary study by \cite{DBLP:journals/paccmod/CiacciaM24} involved a small survey where 44 graduate students evaluated the selection of top basketball players; in that survey, results showed that 66\% of participants favored the directional query outcomes. However, given the constraints of this small, homogeneous group and the narrow topic of the survey, further research is essential to more accurately understand user preferences for directional queries, particularly in varied fields and with diverse user demographics.

While linear top-k queries are user-friendly in certain aspects—allowing users to control the cardinality of the result set and assign weights to attributes—they exhibit notable limitations. Many users prefer results that excel across multiple attributes rather than being ranked solely by a composite score. Traditional Skyline queries \cite{boerzsoenyi2001skyline} provide all non-dominated results but lack flexibility for users to express preferences, resulting in a one-size-fits-all approach. Additionally, Skyline queries do not allow users to limit the result size, often producing excessively large outputs. 

To address this, Regret-Minimizing Sets (RMS) \cite{Computingk-regretminimizing,Regret-Minimizing} were proposed to minimize the maximum regret ratio — the relative difference between the utility of the optimal result in the full dataset and that in the selected subset. By focusing on this metric, RMS ensures that the chosen subset closely approximates the ideal set of results, offering a more compact yet meaningful representation of the dataset. Extensions like non-linear utility functions \cite{nonlinearutilities} further reduce regret and output size. However, RMS methods, fail to support personalization, making them less effective in catering to diverse user needs. Building on RMS, Interactive Regret Minimization approaches \cite{Interactiveregret,StronglyTruthfulInteractive, Sorting-basedInteractiveRegret} enhance personalization by involving users directly in the selection process. While this leads to more tailored outcomes, the requirement for active user participation may unintentionally increase the cognitive load and complexity for users.

To overcome the cardinality control limitations of traditional Skyline queries, researchers have introduced several variants, such as Skyline Frequency \cite{OnHighDimensionalSkylines}, Strong Skyline Points \cite{strongskylinepoints}, Branch-and-Bound Skyline (BBS) \cite{BBS}, and Top-k Dominating Queries \cite{EfficientProcessing,top-kdominatingqueries}, among others
\cite{24,35,34}, which are concluded in the work \cite{DBLP:journals/paccmod/CiacciaM24}.

In parallel, significant work has been done to merge the benefits of Top-k and Skyline queries into a cohesive solution \cite{k-Skyband,Reconciling,FlexibleScoreAggregation,RestrictedSkylines,FlexibleScoreAggregation(ExtendedAbstract),2021marrying,Flexible_Skylines}. In particular, Flexible Skylines \cite{Flexible_Skylines} introduced the concept of \( F \)-dominance, this approach allows users to specify constraints on weight vectors, prioritizing certain attributes (e.g., price, quality, distance) over others. The constraints form a family of scoring functions \( F \), where an object \( t \) is said to \( F \)-dominate another object \( s \) if it scores better in all functions within \( F \). This means that \( t \) aligns more closely with user preferences across all possible configurations. In \cite{2021marrying}, the authors introduced the ORD and ORU operators after evaluating the limitations of traditional top-k and skyline queries. These operators are designed to combine the strengths of both methods, enabling personalized, size-controllable output with flexible preference input that allows users to adjust preferences dynamically. The ORD operator employs adaptive dominance to control the output size efficiently, making it suitable for applications requiring subsecond responses. On the other hand, ORU uses a utility-based ranking approach, which, while slower, shows potential for enhancement through parallelization.

Although these approaches have been validated through experiments and real-world datasets, it is noteworthy that no user surveys were conducted to evaluate user satisfaction or preferences regarding these methods. This stands in contrast to the survey-based evaluation of directional queries, where direct user feedback provided valuable insights into the approach's practical relevance. Incorporating similar evaluations for these methods could help bridge the gap between algorithmic performance and real-world user experience, offering a more comprehensive understanding of their impact.

\section{Methodology}
To achieve the objective of understanding real-world user preferences between linear and directional top-k queries, and to explore the scenarios in which users favor one approach over the other, a structured questionnaire was designed.  Among the various applications of top-k queries, e-commerce and web search stand out as the most widely used, impactful, and influential scenarios. Therefore, this study aligns its dataset selection and questionnaire focus with these two prevalent use cases.
Two sets of realistic datasets were utilized to simulate query results, allowing participants to express their preferences based on their experiences.

This section provides an overview of the questionnaire, covering its general design, the hypotheses of the research question, the theoretical foundations used for its creation, the datasets employed to generate the displayed graphs, the details of the questions, and the data collection and analysis methodology.

\subsection{General design}

This study expands on the initial survey conducted in \cite{DBLP:journals/paccmod/CiacciaM24} to assess real-world user preferences for the results of Directional top-k Query and Linear Top-k Query. To minimize the cognitive load on participants \cite{CognitiveLoad}, and to ensure unbiased results, the survey does not require participants to understand the underlying computational differences between the two query methods. Instead, the results are presented through scatter plots of actual datasets, displaying their differences clearly.

As described in Section 1.1.1, both methods rank objects in ascending order, meaning lower scores result in higher rankings. This is represented graphically as points closer to the origin (0, ... 0) achieving better rankings. To further reduce cognitive burden, especially given the diverse age and cultural backgrounds of participants, the study simplifies the datasets to only two attributes which were selected by the researcher. These attributes are plotted as two-dimensional scatter plots, where both attributes are assigned equal weights of 50\%. Consequently, the preference line (PL) for participants is represented as the diagonal from (0,0) to (1,1). Generally, Linear top-k query results tend to favor objects with extreme values in individual attributes, leading to a more scattered distribution closer to the axes. In contrast, directional query results are more balanced across attributes, forming a distribution along the diagonal and clustering near the preference line.

Additionally, since real datasets are used for queries and rankings, the survey incorporates self-reporting questions to access participants' familiarity with the topics and assess whether their choices were influenced by personal experiences. This approach helps researchers understand whether participants were consciously aware of external factors affecting their decisions. To further mitigate topic-specific bias, the study includes two distinct topics in the survey. By doing so, it compares the applicability of directional query across different scenarios and ensures that conclusions are not overly influenced by a single topic.

\newtheorem{hypothesis}{Hypothesis}

\subsection{Hypotheses}
To structure this investigation, three key hypotheses are proposed, each designed to address a distinct aspect of user preference and behavior. These hypotheses will guide the analysis and help validate the research objectives outlined in this study.

\begin{hypothesis}
There is a statistically significant difference in user preference for the results of linear top-k query and directional top-k query.
\end{hypothesis}

This hypothesis seeks to investigate whether the differences in ranking logic between the two methods lead to meaningful variations in user preference. Linear top-k queries prioritize weighted sums, while directional top-k queries focus on balancing multiple attributes, providing results that better reflect trade-offs. By analyzing user feedback, this hypothesis aims to determine whether these theoretical distinctions translate into practical relevance for users. Validating this hypothesis will help confirm whether directional queries' advantage in addressing attribute balance resonates with user expectations and decision-making.

\begin{hypothesis} The type of dataset topic influences the level of significance of difference between 2 type of queries.
\end{hypothesis}

This hypothesis focuses on exploring how the nature of the dataset topic affects user preferences for query methods. For example, topics closely related to everyday life, such as used car data, may be more easily understood and relatable for general users. Consequently, the preference differences between linear and directional query methods could be more pronounced for such datasets. In contrast, for topics with a higher degree of specialization—such as sports performance metrics or industry-specific complex datasets—users may pay less attention to the details of query results, leading to less significant differences in preferences between the two methods. Validating this hypothesis will help identify the role dataset characteristics play in shaping user preferences and provide insights for optimizing query methods for different application scenarios.

\begin{hypothesis}
The extent of knowledge about the dataset topic influences the user preference between 2 type of queries.
\end{hypothesis}

This hypothesis aims to explore how users with prior knowledge of the dataset topic (e.g., football players’ skills or used car market data) tend to make decisions based on their individual needs and preferences, rather than solely comparing the technical features of linear and directional query methods. For instance, users familiar with the dataset topic may have a clearer idea of the attributes or objects they prioritize. As a result, their selection process is more personalized and driven by their specific requirements rather than relying on the rankings. In contrast, users with limited knowledge of the topic may depend more on the default ranking logic of the query methods. Validating this hypothesis helps to understand how user background knowledge impacts their decision-making when interacting with query methods.

\subsection{Datasets and Questionnaire design}

\subsubsection{Datasets}

The two real-world datasets utilized for the questionnaire were sourced from the website \textit{Kaggle}, a renowned platform for machine learning and data science enthusiasts. Kaggle serves as a dynamic community where beginners and professionals can learn, exchange ideas, and participate in competitions. One of its most popular features is its extensive repository of datasets. It provides access to a repository containing over 50,000 real-world datasets from various domains, enabling comprehensive data analysis and exploration\cite{kaggleGuide2023}. This resource was utilized in this study to ensure the use of authentic and diverse datasets.

\subsubsection*{Used Cars Dataset}

\noindent Source: cars.com

\noindent Total number of tuples: 4009

\noindent Total number of attributes: 12

\noindent Selected attributes: Mileage and Price

The Used Cars dataset originates from cars.com, a leading American automobile classifieds website known for its prominence in the car transaction market\cite{cars_about}. The dataset, updated in 2023, consists of unique entries that combine attributes of brand, model, manufacturing year, and color. In this study, the attributes of mileage and price were selected for analysis. Mileage, a measure of the distance traveled by a car, serves as an important indicator of its usage and potential maintenance needs. Price, reflecting the car's 2023 market value, is another critical factor influencing consumer decision-making. These two attributes were chosen because they represent the most relevant dimensions that consumers typically weigh when purchasing cars online.

The topic choice of cars for this study is rooted in the High-Involvement Decision-Making Theory\cite{BuyerBehavior}, which contrasts routine decisions for low-cost goods with significant, high-stakes purchases. Buying a car, unlike routine purchases such as shampoo or a cup of coffee, involves considerable time, effort, and emotional engagement, as it is tied to the buyer's self-perception and includes financial and psychological risks. In these scenarios, consumers are motivated to thoroughly evaluate options and consider trade-offs. Top-k query methodologies are particularly valuable in this context, as they assist in ranking alternatives, and supporting consumers in making well-informed decisions for purchases that matter deeply to them.

\subsubsection*{Football Players Dataset}

\noindent Source: SoFIFA.com \\
\noindent Total number of tuples: 3012 \\
\noindent Total number of attributes: 64 \\
\noindent Selected attributes: total skill score and total defending score

To enrich the research and further explore user preferences for query results in different contexts, this study includes a second dataset focused on football players, which belongs to the domain of personal interests, reflecting an individual's engagement with sports and hobbies rather than practical purchasing decisions.  The football players dataset originates from SoFIFA.com, a leading and innovative online FIFA/EAFC series career mode database. This dataset serves as a reliable resource, bridging in-game data with real-world football statistics\cite{sportmonks_sofifa}. Retrieved in 2024, the dataset reflects the latest updates, containing 64 columns, 55 of which are attribute scores used to evaluate various aspects of players' football abilities.

For this study, total skill and total defending were selected as the two main attributes. Total skill encompasses metrics such as Dribbling, Curve, Free Kick Accuracy, Long Passing, and Ball Control, representing offensive capabilities. Total defending includes attributes such as Defensive Awareness, Standing Tackle, and Sliding Tackle, highlighting defensive performance. These two dimensions were chosen because they comprehensively represent a player's performance in both offensive and defensive scenarios, providing a holistic view of their overall ability. This approach provides users with a more integrated and persuasive standard for evaluating players, offering a complete perspective on their overall effectiveness in key match situations.

\subsubsection{Data cleaning and preprocessing}

Before implementing the query results using these datasets, it is crucial to conduct a thorough data cleaning process to ensure the reliability of the final outputs. This step is essential for identifying inconsistencies, eliminating errors, and minimizing the risk of mistakes that could lead to rework or inaccuracies during later stages.

To carry out this process effectively, \textit{Python} and \textit{Jupyter Notebook} are utilized due to their user-friendly interface and robust capabilities for data processing and visualization. Specifically, the \textit{pandas} library plays a central role in this workflow. It allows seamless data manipulation, such as handling missing values, detecting duplicates, and performing transformations. Additionally, Jupyter Notebook provides an interactive environment where code, visualizations, and documentation can be combined. 

\subsubsection*{Used Cars Dataset}

As previously mentioned, the Used Cars dataset uniquely identifies each car through a combination of its brand, model, manufacturing year, and external color, as illustrated in Fig. 3. To ensure clarity and precision in handling the data, we created a unique index by combining these four attributes. This index simplifies data representation and ensures that each car is uniquely identified within the dataset. 
\begin{figure}
    \centering
    \includegraphics[width=1\linewidth]{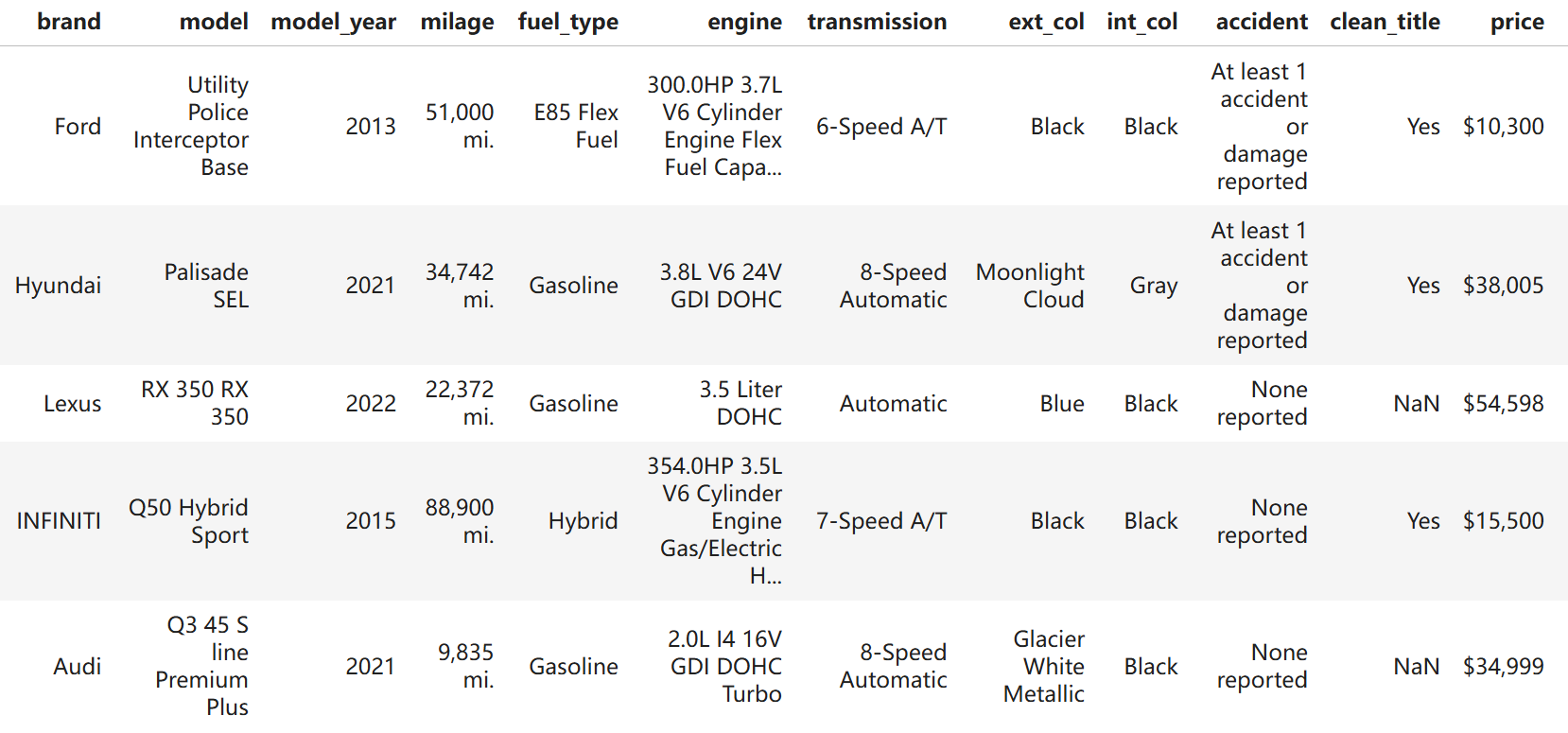}
    \caption{Used Cars dataset overview}
    \label{fig:enter-label}
\end{figure}
In the cleaning process, we identified that there is no null values in the dataset, but we addressed several formatting inconsistencies. The mileage attribute, for example, included the unit "mi" within the values, which was converted into a pure numeric format to facilitate calculations. Similarly, the price attribute included currency symbols, which were removed and converted into numeric values for further analysis. After completing these initial cleaning and transformation processes, the dataset structure is displayed in Fig. 4. 
\begin{figure}
    \centering
    \includegraphics[width=1\linewidth]{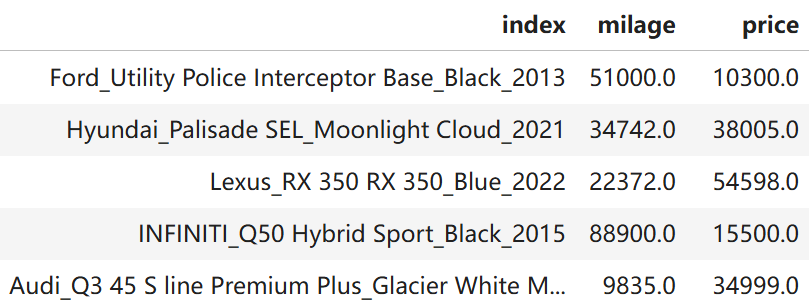}
    \caption{Head of Used Cars dataset after cleaning}
    \label{fig:enter-label}
\end{figure}
To further refine the data, we performed outlier detection to ensure accurate rankings. By analyzing boxplots for the mileage and price attributes (shown in Fig. 5), we determined that only cars priced below \$100,000 and with mileage under 200,000 miles would be retained. This additional step was necessary to eliminate the influence of extreme values on the ranking results, enhancing the reliability of subsequent analyses.

\begin{figure}
    \centering
    \begin{subfigure}{0.48\textwidth} 
        \includegraphics[width=\linewidth]{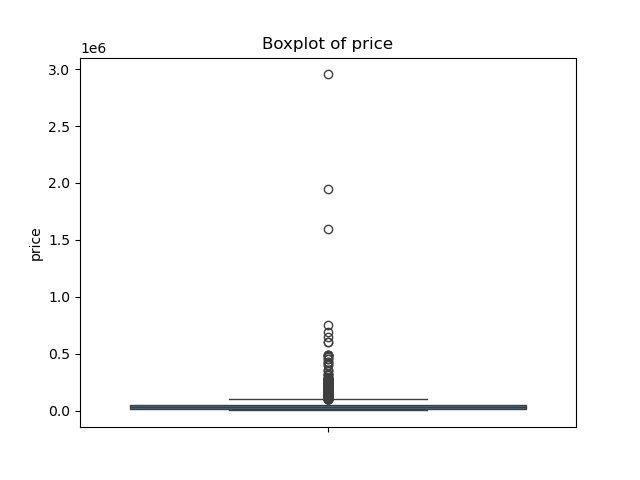}
        \caption{Boxplot of Price} 
        \label{fig:boxplot_price}
    \end{subfigure}
    \hfill 
    \begin{subfigure}{0.48\textwidth}
        \includegraphics[width=\linewidth]{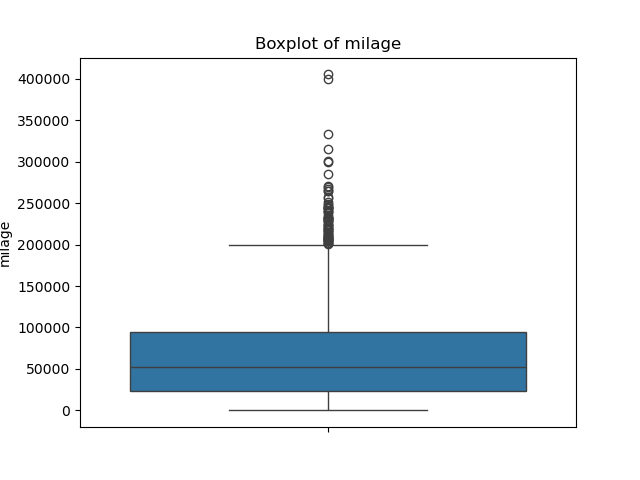}
        \caption{Boxplot of Mileage}
        \label{fig:boxplot_mileage}
    \end{subfigure}
    \caption{Boxplots of Price and Milage in Used Cars dataset}
    \label{fig:boxplots}
\end{figure}

\subsubsection*{Football Players Dataset}

Fig. 6 provides an overview of the football dataset. Unlike the used cars dataset, no formatting transformations were required for the values. However, the dataset contained null values, particularly in the total skills and total defending attributes. To maintain data quality, we excluded all tuples with null values in these fields, reducing the dataset to 2,399 unique football players suitable for analysis. The resulting dataset, cleaned and prepared for queries, is illustrated in Fig. 7.

\begin{figure}
    \centering
    \includegraphics[width=1\linewidth]{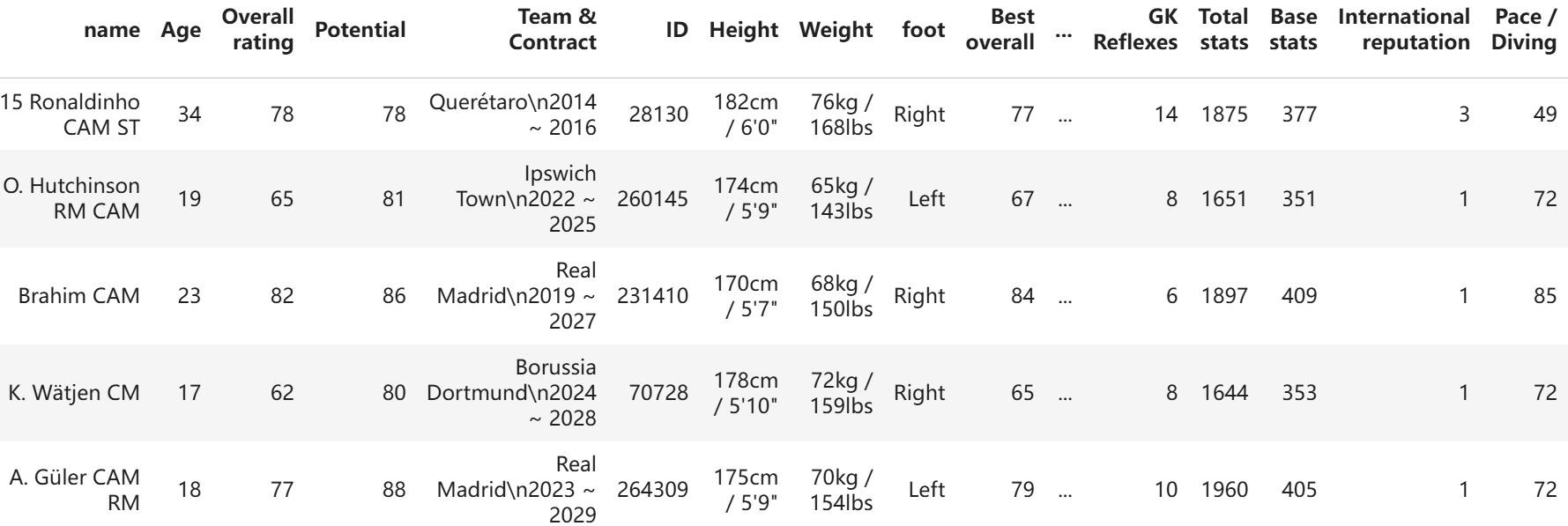}
    \caption{Football Players dataset overview}
    \label{fig:enter-label}
\end{figure}

\begin{figure}
    \centering
    \includegraphics[width=0.75\linewidth]{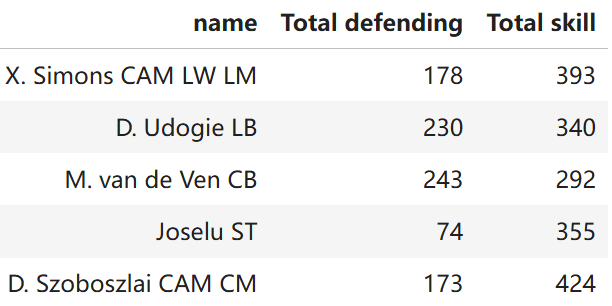}
    \caption{Head of Football Players dataset after cleaning}
    \label{fig:enter-label}
\end{figure}

Similar to the used cars dataset, the football players dataset also required the removal of outliers to prevent them from skewing the final ranking results. After analyzing the boxplots for total skill and total defending attributes in Fig. 8, we filtered the dataset to include only football players with total skill scores exceeding 160.

\begin{figure}
    \centering
    \begin{subfigure}{0.48\textwidth} 
        \includegraphics[width=\linewidth]{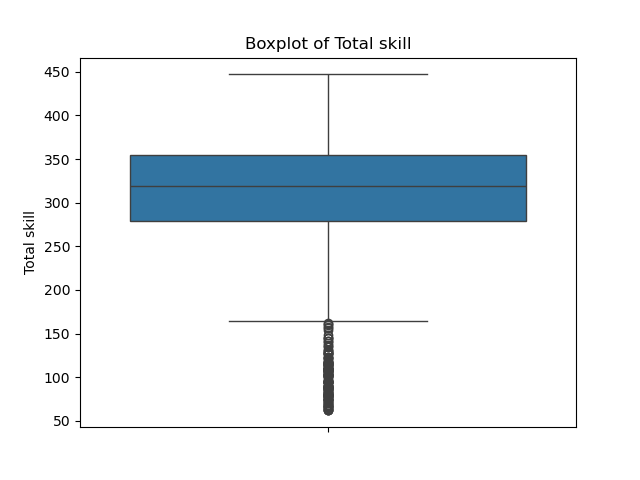}
        \caption{Boxplot of Total skill} 
        \label{fig:boxplot_total_skill}
    \end{subfigure}
    \hfill 
    \begin{subfigure}{0.48\textwidth}
        \includegraphics[width=\linewidth]{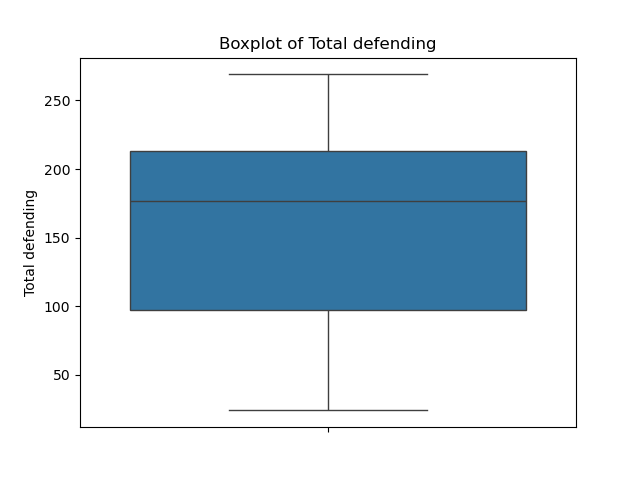}
        \caption{Boxplot of Total defending}
        \label{fig:boxplot_total_defending}
    \end{subfigure}
    \caption{Boxplots of Skill and Defending in Football Players dataset}
    \label{fig:boxplots}
\end{figure}

\subsubsection*{Data Normalization}

After completing the data cleaning process, a crucial common step for both datasets is data normalization using Min-Max normalization \cite{MinMaxNormalization2021}, which scales the selected attributes in each dataset to a range of [0,1]. It's a vital step because it will ensure that both attributes contribute equally to the scoring and ranking processes. Before normalization, attributes with larger numerical ranges might influence the results, skewing the rankings and making comparisons less meaningful. 

\subsubsection{Questionnaire design}

At the beginning of the questionnaire, participants are provided with a concise introduction to the study. This includes a brief explanation of linear top-k queries and directional top-k queries, as well as an overview of the practical applications of top-k queries in real-world scenarios. Additionally, the key difference between these two methods is highlighted: directional top-k queries are designed to produce results that are more balanced between the two selected attributes compared to linear top-k queries. This introduction ensures participants have a foundational understanding of the two approaches before delving into the questionnaire. Importantly, we clarify that respondents do not need any specialized knowledge about used cars or football players to answer the questions, as the questionnaire is designed to be inclusive and accessible to a broad audience, not solely experts. Furthermore, we also include a Confidentiality Assurance statement at the beginning, clearly informing participants that their responses will remain strictly confidential and will only be used for academic research purposes. No personal information will be disclosed under any circumstances.

The main body of questionnaire consists of three sections: the used cars topic, the football players topic, and the personal information section. The personal information section includes questions about the participants' age and gender. In the first two sections, scatter plots are used to present the query results from linear top-k and directional top-k queries separately. As shown in Fig. 9, different colors are utilized to distinguish the methods: blue for linear top-k query results, green for directional query results, and yellow for overlapping results. In the used cars dataset, each scatter point is annotated with the unique index of the car, which includes its brand, model, manufacturing year, and color, as well as its price and mileage, providing participants with detailed information. Similarly, in the football players dataset, scatter points are labeled with the player’s name along with their total skill and total defending scores. For yellow points representing overlapping results, no text annotations are included to avoid unnecessary emphasis, allowing participants to focus solely on the differences between the two query results.

\begin{figure}
    \centering
    \begin{subfigure}{0.7\textwidth} 
        \includegraphics[width=\linewidth]{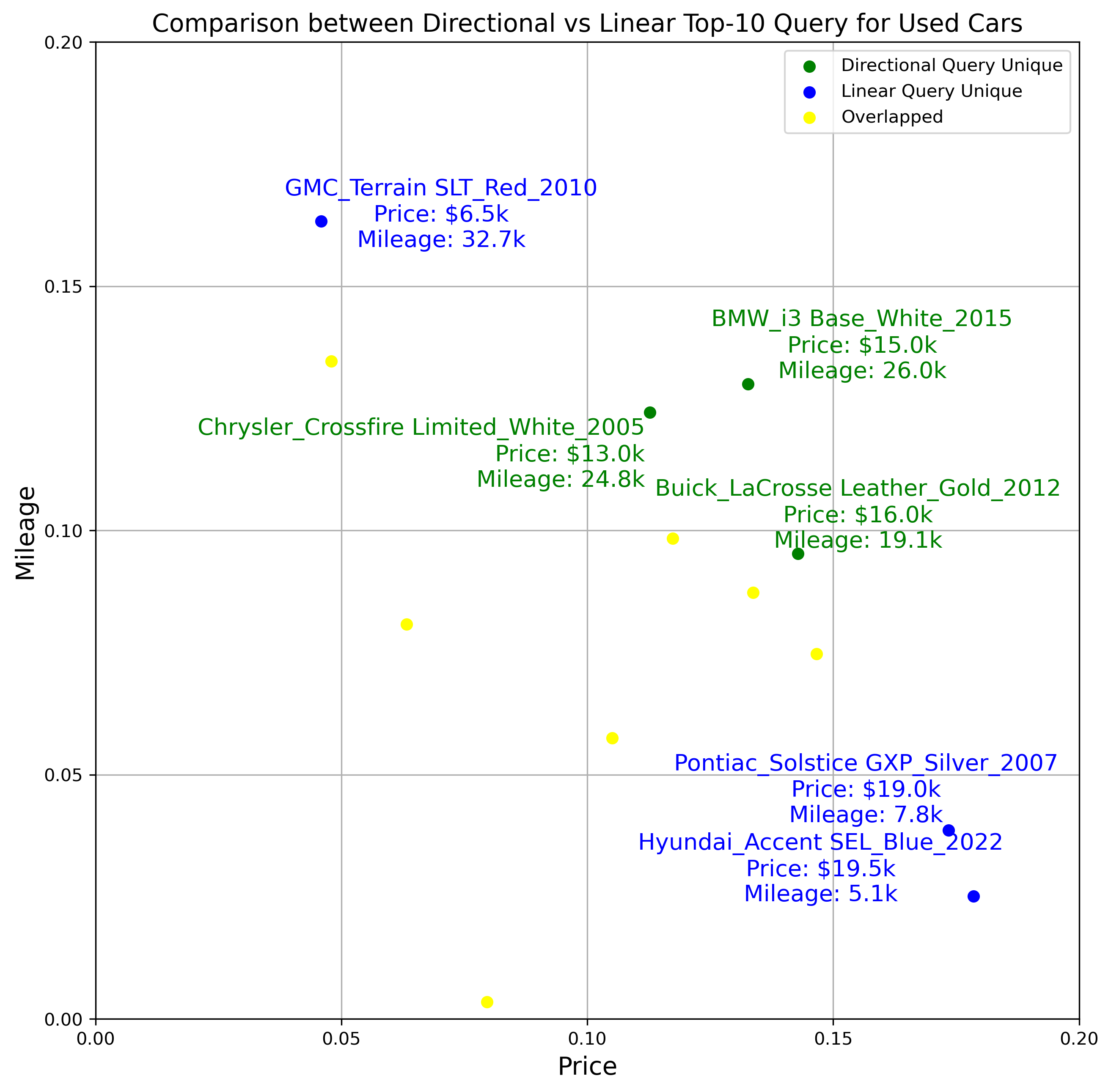}
        \caption{Used Cars query results} 
        \label{fig:Used Cars query results}
    \end{subfigure}
    \hfill 
    \begin{subfigure}{0.7\textwidth}
        \includegraphics[width=\linewidth]{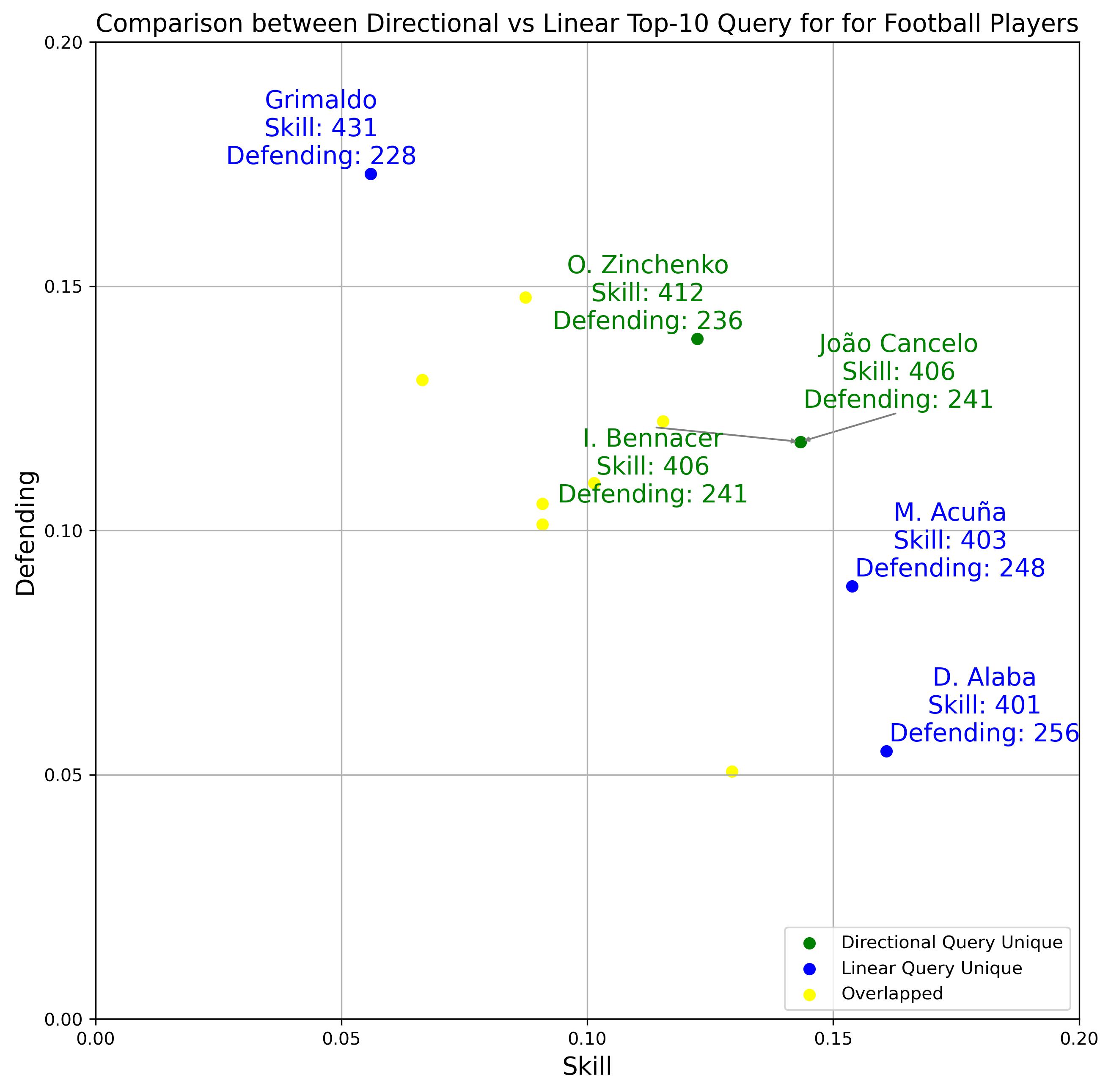}
        \caption{Football Players query results}
        \label{fig:Football Players query results}
    \end{subfigure}
    \caption{Query results of Used Cars and Football Players}
    \label{fig:boxplots}
\end{figure}

In addition to selecting between linear top-k query and directional top-k query, participants are also asked to respond to self-report questions. These questions aim to assess the participants' level of knowledge on the topic and to allow them to self-evaluate the extent to which their choices were influenced by their prior knowledge. Each part of the questionnaire includes only three straightforward questions:

\subsubsection*{Used Cars dataset questions}
\begin{itemize}
    \item In our analysis of over 4,000 second-hand cars listed on cars.com, each vehicle has been evaluated based on price and mileage. Both attributes are scored on a scale from 0 (best) to 1 (worst), both attributes are considered equally important.

    The most cost-effective cars are those that appear closest to the bottom-left corner of the chart. This positioning indicates lower prices and lower mileage, making these vehicles the most desirable in terms of value.

    Which results would you prefer to have when you query online about 'the top 10 most cost-effective second-hand cars'?
    \item Have you purchased a used car recently (within the last three years)?
    \item How would you describe your understanding of the used car market?
    \item Do you think your knowledge of used cars influenced your choice of query results above?
\end{itemize}

\subsubsection*{Football Players dataset questions}

\begin{itemize}
    \item In our analysis of over 3,000 football players, each individual’s skill (including Dribbling, Curve, Free Kick Accuracy, Long passing and Ball control) and defending (including Defensive awareness, Standing tackle and Sliding tackle) capabilities are evaluated. Ratings are assigned from 0 (best) to 1 (worst) for both attributes, both attributes are considered equally important. 

    The optimal performers are those located closest to the bottom-left corner of the scatter plot. This position signifies superior skill and defensive abilities, making these players the most proficient.

    Which results would you prefer to have when you query online about 'the top 10 best football players'? 
    \item How often do you watch football matches?
    \item How would you describe your knowledge of football?
    \item Do you think your knowledge of or interest in football influenced your choice of query results above?
\end{itemize}

\subsection{Data collection and analysis}

\subsubsection{Data collection}

Given that the survey focused on the top-k usage scenarios in e-commerce and web searches, participation was open to individuals of all ages, genders, and levels of prior knowledge, ensuring inclusiveness across diverse demographics. Sampling methods can be broadly categorized into probability and non-probability sampling. Probability sampling is designed to ensure the representativeness of the sample and generalizability of the results to the target population. In contrast, non-probability sampling involves methods where the probability of selecting a subject is unknown, often resulting in selection bias in the study \cite{acharya2013sampling,probability}.

 This study employed non-probability sampling methods, specifically convenience sampling and snowball sampling. 
 In Convenience sampling, it involved distributing the questionnaire through online platforms, such as social media, group chats, and public forums, to reach a broad audience easily. Meanwhile, snowball sampling involved leveraging initial respondents to share the questionnaire within their personal and professional networks. This approach created a ripple effect, as each new participant was encouraged to further disseminate the survey, resulting in a chain reaction that significantly expanded the reach and diversity of the participant pool.

\subsubsection{Data analysis}

Data analysis was performed using SPSS Statistics for Windows, version 29.0 (IBM Corp). Descriptive statistics were initially employed to provide an overall understanding of the dataset. For the three hypotheses, query preferences were used as the dependent variable, with other objective factors serving as independent variables. Binomial tests and McNemar’s cross-tabulation analyses were conducted to ensure robust cross-validation of findings. Additionally, binary logistic regression and chi-square tests were utilized to conduct detailed stratified analyses. Statistical significance was defined as \(P \leq 0.05\)  for all tests.

\section{Results and discussion}
This section presents the findings from the user preference survey, focusing on the comparison between linear and directional top-k queries. 
This part will focus on the answer to the Hypotheses that brought out in the previous part in 3.1.1, each hypothesis is evaluated to determine whether the observed data supports or refutes it.

\subsection{Descriptive Results}
\subsubsection{Participant demographics}
A total of 106 individuals participated in the survey, as the inclusion criteria did not impose any restrictions, such as those based on age or gender. Therefore, all responses were included in the analysis. The average time taken to complete the survey was 3 minutes and 33 seconds. Among them, 48\% (51/106) identified as women, 42\% (45/106) as men, 3\% as non-binary, and 7\% preferred not to disclose their gender. The mean age of the participants was 27.78 years old, with a standard deviation of 6.22.  Notably, 66\% of the respondents were between 25 and 34 years old. Additional participant demographics details are presented in Table 1.

\begin{table}[ht]
\centering
\begin{tabular}{@{}ll@{}}
\toprule
\textbf{Question}        & \textbf{Participants, n(\%)}                           \\ \midrule
\textbf{Gender (n=106)}     &                                     \\
\hspace{1em}Male          &  \hspace{1em}45 (42\%)                       \\
\hspace{1em}Female        &   \hspace{1em}51 (48\%) 
    \\ 
\hspace{1em}Non-binary          &   \hspace{1em}3 (3\%)                                 \\
\hspace{1em}Prefer not to say    & \hspace{1em}7 (7\%)                                 \\
\midrule
\textbf{Age in years (n=106)} &                               \\
\hspace{1em}18-24         &       \hspace{1em}32 (30\%)                             \\
\hspace{1em}25-34         &    \hspace{1em}70 (66\%)                                 \\
\hspace{1em}35-44         &    \hspace{1em}0                                 \\
\hspace{1em}45-54         &     \hspace{1em}3 (3\%)                                \\
\hspace{1em}55-64         &      \hspace{1em}1 (1\%)                              \\  \bottomrule
\end{tabular}
\caption{Participant demographics}
\label{tab:Participant demographics}
\end{table}

\subsubsection{User preference for query results in general}

Participants in the study were presented with two scenarios: selecting their preferred query method under the topics of used cars and football players. These scenarios aimed to compare user preferences between directional and linear top-k query results. In the used cars dataset, 70\% of participants (74 individuals) preferred the directional query method over the linear top-k method. Conversely, in the football players section, preferences shifted significantly. Only 44 participants (42\%) favored directional queries, while the majority, 58\% (62 participants), indicated a preference for the traditional linear top-k query method. The bar chart in Figure 10 provides a visual representation of these findings, highlighting the differing preferences between the two topics. 

\begin{figure}
    \centering
    \includegraphics[width=0.75\linewidth]{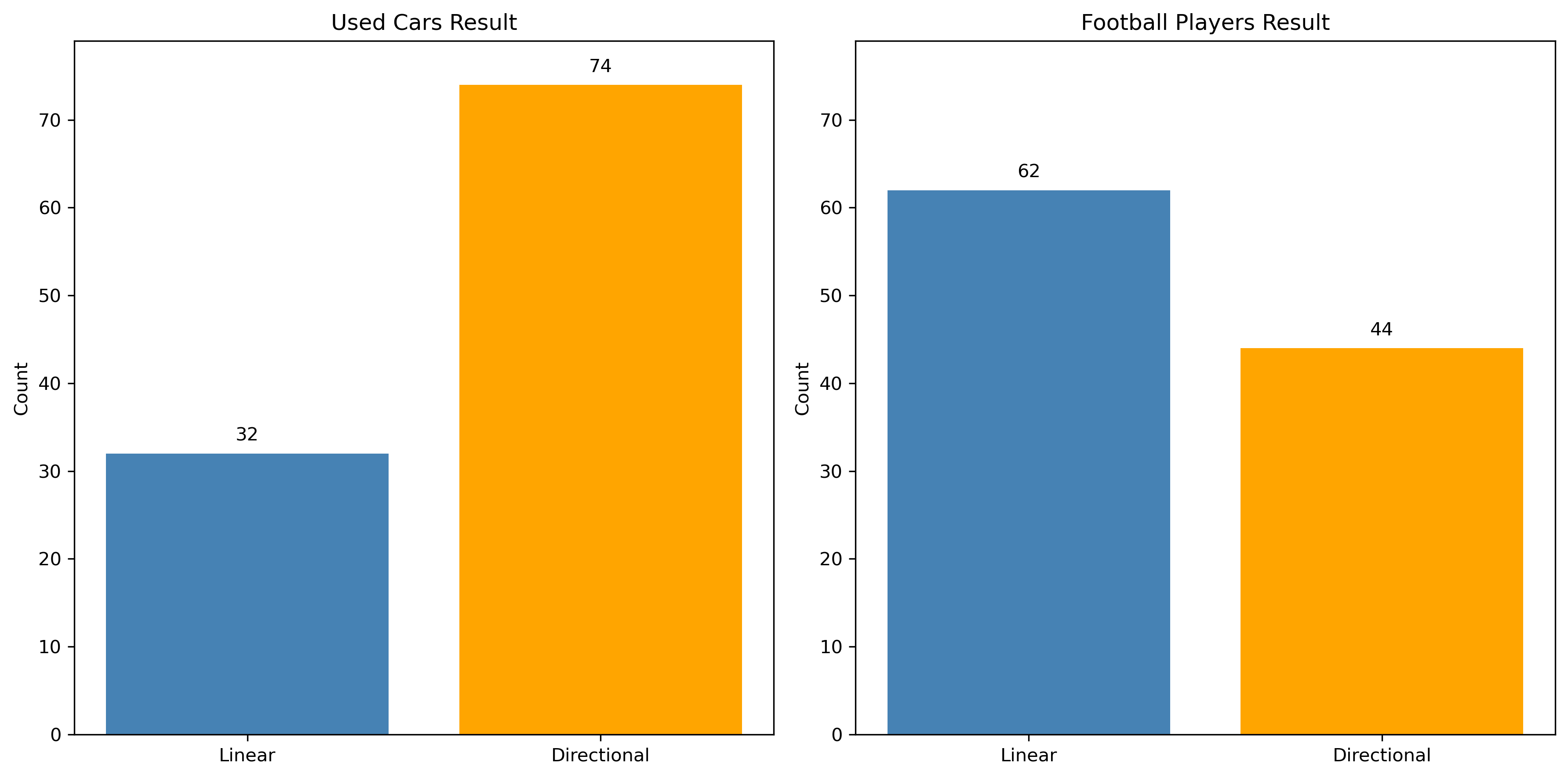}
    \caption{User preference for query results}
    \label{fig:User preference for query results}
\end{figure}

\subsubsection{Users' knowledge and experience for topics in general}

\begin{table}[ht]
\centering
\begin{tabular}{@{}p{5cm}llr@{}}
\toprule
\textbf{Question} & \textbf{Response} & \textbf{Count (n, \%)} \\ \midrule
\multirow{2}{=}{\textbf{Have you purchased a used car recently?}} 
  & Yes & 10 (9\%) \\ 
  & No & 96 (91\%) \\ \midrule
\multirow{4}{=}{\textbf{How would you describe your understanding of the used car market?}} 
  & Very good & 7 (7\%) \\ 
  & Good & 32 (30\%) \\ 
  & Fair & 40 (38\%) \\ 
  & Poor & 27 (25\%) \\ \bottomrule
\end{tabular}
\caption{User knowledge and experience on used cars}
\label{tab:User knowledge and experience on used cars}
\end{table}

\begin{table}[H]
\centering
\begin{tabular}{@{}p{5.5cm}llr@{}}
\toprule
\textbf{Question} & \textbf{Response} & \textbf{Count (n, \%)} \\ \midrule
\multirow{6}{=}{\textbf{How often do you watch} \\ \textbf{football matches?}}
  & Daily & 0 \\ 
  & Weekly & 13 (12\%)\\ 
  & Monthly & 7 (7\%)\\ 
  & Seasonal & 24 (23\%)\\ 
  & Yearly & 38 (36\%)\\ 
  & Never & 24 (23\%)\\ \midrule
\multirow{4}{=}{\textbf{How would you describe your} \\ \textbf{knowledge of football?}}
  & Beginner & 63 (59\%)\\ 
  & Intermediate & 26 (25\%)\\ 
  & Advanced & 13 (12\%)\\ 
  & Expert & 4 (4\%)\\ \bottomrule
\end{tabular}
\caption{User knowledge and experience on football players}
\label{tab: User knowledge and experience on football players}
\end{table}

In the used cars topic, the vast majority of participants (91\%, n=96) had not purchased a used car within the past three years, leaving only 9\% (n=10) who had. Additionally, while 37\% of participants demonstrated good knowledge of the used car market, most reported having limited familiarity.

For the football players topic, only 19\% of participants reported watching matches frequently (weekly or monthly). A significant portion were seasonal or yearly viewers, while 23\% stated they had never watched a football match. Additionally, 59\% of participants self-reported as beginners in football knowledge, with only 16\% identifying as having advanced or expert-level knowledge.

\subsubsection{Self report of the level of knowledge influence}
\begin{figure}[H]
    \centering
    \includegraphics[width=0.75\linewidth]{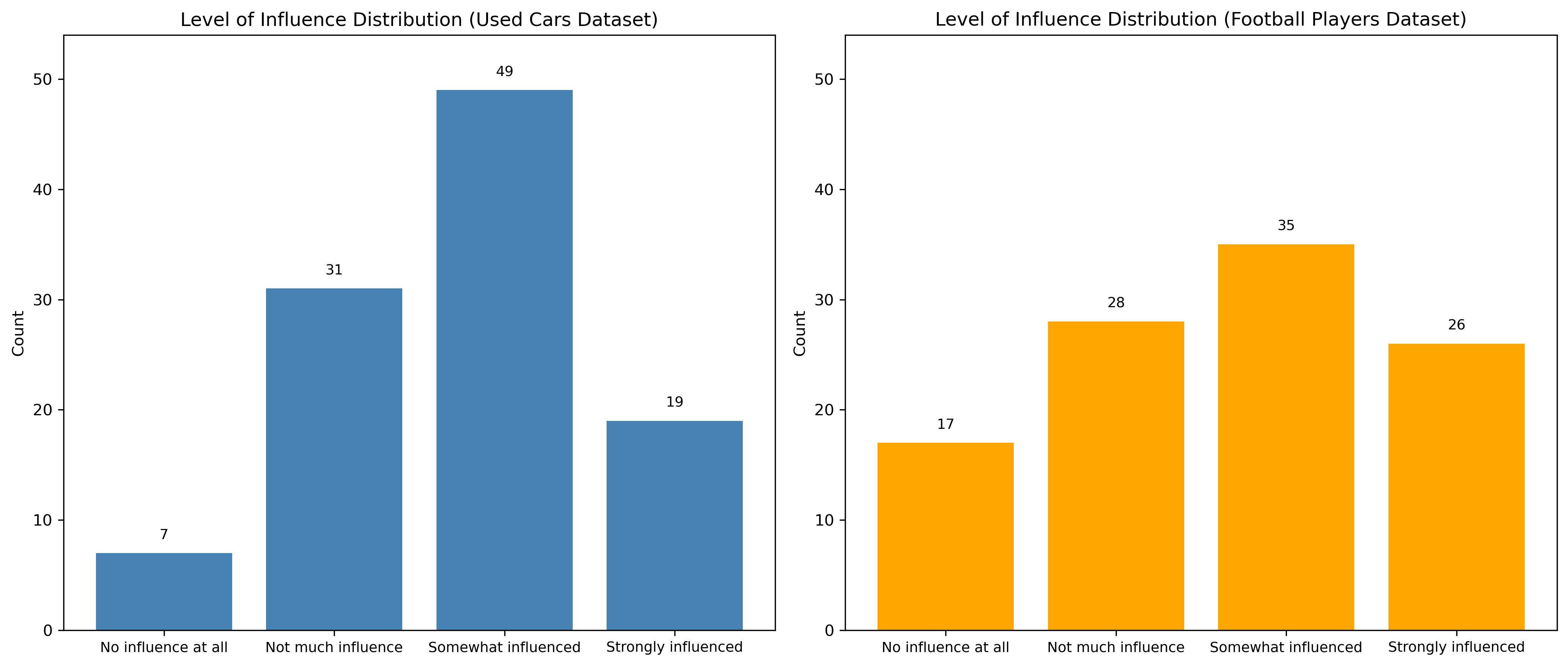}
    \caption{Self-reported influence distribution}
    \label{fig:Influence distribution}
\end{figure}

Figure 11 illustrates the self-reported influence of participants' knowledge and experience on their query preferences across two topics. For the used cars topic, very few participants reported being "not at all influenced" or "strongly influenced," while the majority (80 out of 106) indicated being influenced to some degree. In contrast, for the football players topic, the distribution of responses was more evenly spread across the four levels of influence. Although fewer participants overall reported being influenced compared to the used cars topic, a higher number (26 out of 106) indicated being "strongly influenced."

\subsection{Hypothesis Testing Results}

\subsubsection{Comparison of User Preferences}

\textbf{Hypothesis 1.} There is a statistically significant difference in user preference for the results of linear top-k query and directional top-k query. 

To analyze user preferences, two distinct topics were examined separately using binomial tests. These tests assessed whether users showed a significant preference for one type of query over the other within each topic. In the binomial test, the test hypothesis is 0.5, which means that the proportion of observations in each category is assumed to be equal, that is, there is no obvious preference difference between two different queries.

\begin{figure} [H]
    \centering
    \includegraphics[width=1\linewidth]{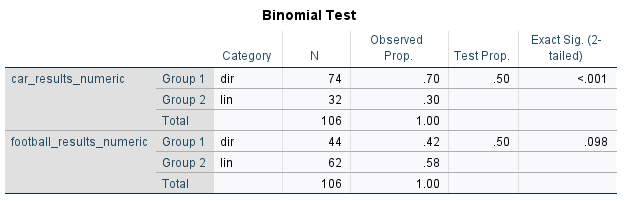}
    \caption{Binomial test of user preference on query results}
    \label{fig:binomial test}
\end{figure}

Figure 12 shows that, for the used cars topic, 70\% of participants preferred directional top-k queries over linear top-k queries. With p<0.001 (significance level < 0.05), this indicates a significant preference for directional queries, partially supporting Hypothesis 1: users exhibit a significant preference for directional queries. For the football players topic, however, 58\% of users favored linear queries, and 42\% preferred directional queries. With a p-value of 0.098, this result suggests that the preference for linear queries does not significantly differ from the expected 50\% proportion, indicating no pronounced preference for either query method. Thus, Hypothesis 1 is supported for the used cars topic but not for the football players topic.

\begin{figure}[ht]
    \centering
    \begin{subfigure}[b]{0.45\textwidth}
        \includegraphics[width=\textwidth]{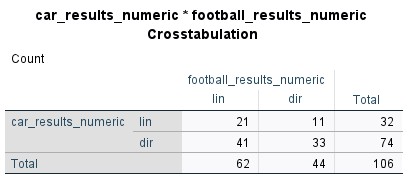}
        \caption{Crosstabulation Results}
        \label{fig:crosstab}
    \end{subfigure}
    \hfill
    \begin{subfigure}[b]{0.45\textwidth}
        \includegraphics[width=\textwidth]{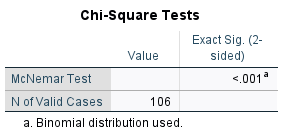}
        \caption{McNemar Test Results}
        \label{fig:mcnemar}
    \end{subfigure}
    \caption{Crosstabulation and McNemar Test}
    \label{fig:McNemar}
\end{figure}

To explore whether individual respondents displayed significantly different preferences across the two topics, McNemar’s test was conducted. As shown in Figure 13, the result (p<0.01) reveals a significant difference in choices between the two topics, indicating that the data cannot be directly aggregated for joint analysis. This finding transitions into the evaluation of Hypothesis 2.

\subsubsection{Cross-Topic Analysis}

\textbf{Hypothesis 2.} The type of dataset topic influences the level of significance of difference between 2 type of queries.

In the previous analysis, the McNemar test confirmed a significant difference between the two topics in terms of user preferences. In this subsection, binary logistic regression is employed to further validate whether the topic serves as a significant factor influencing user choices between these two different datasets. In this regression model, the topic serves as the independent variable, while the dependent variable reflects the users' selection between directional and linear queries. Specifically, the analysis aims to examine whether the topic variable influences users' query choices and to investigate whether additional variables may play a role in shaping their preferences. Figure 14 presents the findings, offering a more detailed understanding of the role of the topic in influencing user preferences.

\begin{figure}[ht]
    \centering
    \begin{subfigure}[b]{0.5\textwidth}
        \includegraphics[width=\textwidth]{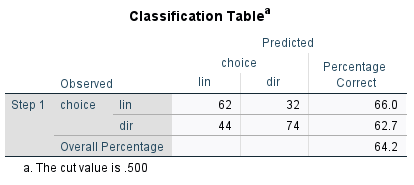}
        \caption{Classification table}
        \label{fig:Classification table}
    \end{subfigure}
    \hfill
    \begin{subfigure}[b]{0.5\textwidth}
        \includegraphics[width=\textwidth]{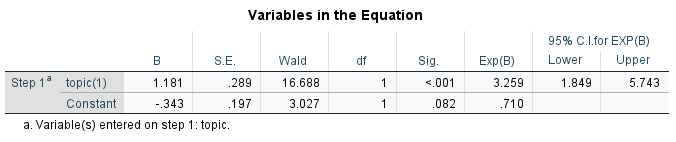}
        \caption{Logistic regression results}
        \label{fig:Logistic regression for cross-topic}
    \end{subfigure}
    \caption{Logistic regression for cross-topic}
    \label{fig:Logistic regression for cross-topic}
\end{figure}

Figure 14(b) shows a p-value of less than 0.001, confirming that the two distinct topics selected in this research exhibit statistically significant differences in their influence on user preferences. The Exp(B) value of 3.259 indicates that when the topic is used cars, users are 3.259 times more likely to choose the directional query (dir) compared to the football players topic. This supports the finding that users exhibit a stronger preference for the directional query when discussing car-related topics, whereas preferences for the football topic are more evenly distributed, resembling random selection.

However, as shown in Figure 14(a), the model achieves an overall prediction accuracy of 64.2\%. While the topic variable accounts for some variation in user preferences, it is insufficient to fully explain these patterns. This suggests the potential influence of additional factors, such as user expertise, interest, or experience, in driving their query choices.

\subsubsection{Knowledge and experience level and User preferences}

\textbf{Hypothesis 3.} The extent of knowledge about the dataset topic influences the user preference
between 2 type of queries.

In the survey, participants were asked two questions for each of the two topics. One question focused on the frequency of real-life involvement with the topic (e.g., whether they had recently purchased a used car), while the other addressed their self-reported knowledge of the topic. This section applies logistic regression with a multivariable approach to analyze how these variables influence user preferences. 

\begin{figure}[ht]
    \centering
    \begin{subfigure}[b]{0.5\textwidth}
        \includegraphics[width=\textwidth]{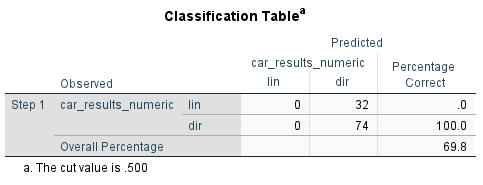}
        \caption{Classification table}
        \label{fig:Classification table}
    \end{subfigure}
    \hfill
    \begin{subfigure}[b]{0.5\textwidth}
        \includegraphics[width=\textwidth]{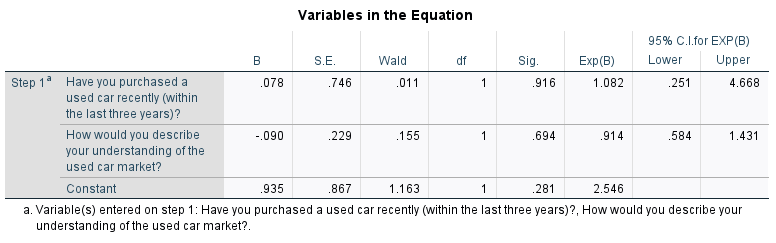}
        \caption{Logistic regression results}
        \label{fig:Logistic regression for used cars knowledge}
    \end{subfigure}
    \caption{Logistic regression for used cars topic}
    \label{fig:Logistic regression for used cars topic}
\end{figure}

\begin{figure}[ht]
    \centering
    \begin{subfigure}[b]{0.5\textwidth}
        \includegraphics[width=\textwidth]{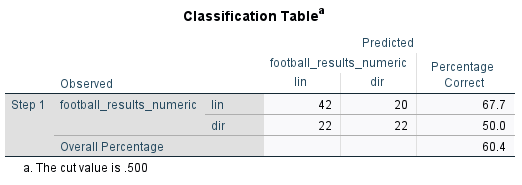}
        \caption{Classification table}
        \label{fig:Classification table}
    \end{subfigure}
    \hfill
    \begin{subfigure}[b]{0.5\textwidth}
        \includegraphics[width=\textwidth]{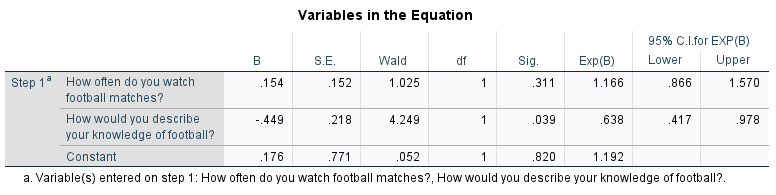}
        \caption{Logistic regression results}
        \label{fig:Logistic regression for football knowledge}
    \end{subfigure}
    \caption{Logistic regression for football topic}
    \label{fig:Logistic regression for football topic}
\end{figure}

As shown in Fig.15(b), the p-values for both questions regrading used cars are greater than 0.05, indicating no significant impact of participants' knowledge and experience on their preference for query types. In contrast, Fig.16(b) reveals a significant p-value of 0.039 (<0.05) for football knowledge, suggesting that football knowledge does influence user preferences, while viewing frequency has negligible influence. Despite these findings, the logistic regression models’ limited classification accuracy (69.8\% for used cars and 60.4\% for football) indicates the presence of other influential factors. Considering the small sample size and to gain a clearer understanding of the role of football knowledge, we performed a chi-square test focusing on different knowledge categories. To minimize the effect of sample size constraints, we grouped the knowledge levels into two categories: "Lower Knowledge" (combining Beginner and Intermediate) and "Higher Knowledge" (combining Advanced and Expert). The results of this restructured analysis are displayed in Fig.17.

\begin{figure}[ht]
    \centering
    \begin{subfigure}[b]{0.5\textwidth}
        \includegraphics[width=\textwidth]{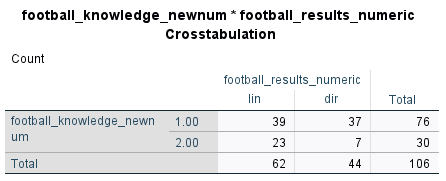}
        \caption{Crosstabulation Result}
        \label{fig:crosstab}
    \end{subfigure}
    \hfill
    \begin{subfigure}[b]{0.5\textwidth}
        \includegraphics[width=\textwidth]{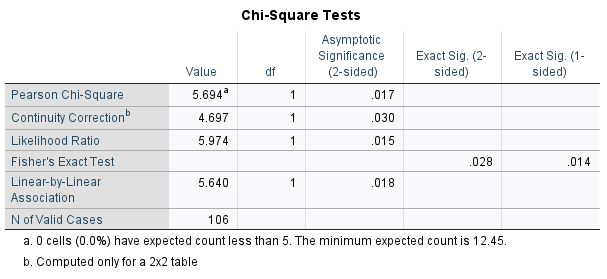}
        \caption{Chi-Square Test Results}
        \label{fig:Chi-Square Test}
    \end{subfigure}
    \caption{Crosstabulation and Chi-Square Test}
    \label{fig:Chi-Square}
\end{figure}

The chi-square test results reveal a significant relationship (p < 0.05) between football knowledge and query preferences. Participants with lower football knowledge (Beginner and Intermediate) displayed a more evenly distributed selection pattern between the two query types. Conversely, those with higher football knowledge (Advanced and Expert) exhibited a strong preference for the linear top-k query, with 23 of 30 respondents in this group favoring it.

\subsection{The effect of demographics on user preferences}

In this section, we broaden the scope of our analysis to examine additional factors, such as gender and age, that may influence user preferences beyond the initial hypotheses. Fig.18 illustrates that in the context of the used cars topic, neither gender nor age had a significant effect. However, for the football topic, age demonstrated statistical significance, with a p-value of 0.048 (<0.05), indicating a meaningful relationship between age and query preferences. To gain deeper insights into how age influences user preferences within the football topic, we conducted a chi-square test using a crosstab analysis, focusing solely on age and participants' choices in this topic.

\begin{figure}[H]
    \centering
    \begin{subfigure}[b]{0.5\textwidth}
        \includegraphics[width=\textwidth]{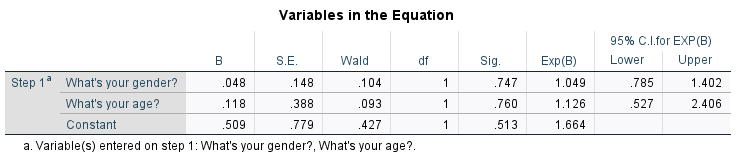}
        \caption{Influence on used cars topic}
        \label{fig:used cars topic}
    \end{subfigure}
    \hfill
    \begin{subfigure}[b]{0.5\textwidth}
        \includegraphics[width=\textwidth]{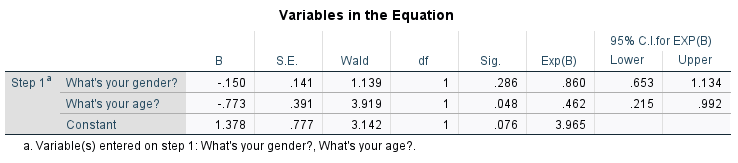}
        \caption{Influence on football players topic}
        \label{fig:football players topic}
    \end{subfigure}
    \caption{Age and gender factor on user preference}
    \label{fig:Age and gender}
\end{figure}

\begin{figure}[H]
    \centering
    \begin{subfigure}[b]{0.5\textwidth}
        \includegraphics[width=\textwidth]{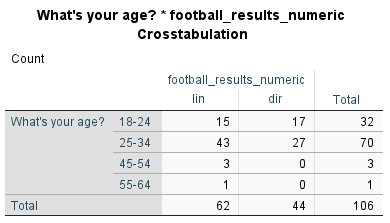}
        \caption{Crosstabulation Result}
        \label{fig:Crosstabulation Result}
    \end{subfigure}
    \hfill
    \begin{subfigure}[b]{0.5\textwidth}
        \includegraphics[width=\textwidth]{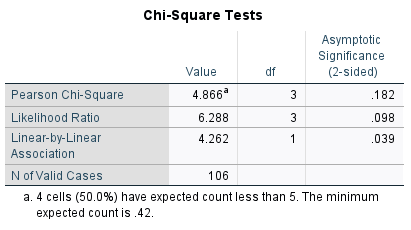}
        \caption{Chi-Square result}
        \label{fig:Chi-Square result}
    \end{subfigure}
    \caption{Chi-Square test for age influence on user preference}
    \label{fig:Chi-Square test for age influence on user preference}
\end{figure}

From the results of the chi-square test, it is evident that the impact of age on participants’ query preferences within the football topic is only marginally significant. Specifically, the p-value for the Linear-by-Linear Association is 0.039, indicating a potential linear relationship between age and the choice of query method.

\subsection{Self-reported and actual influences on user preferences}

In the survey, to better understand how participants' self-perceived knowledge influenced their choices of query results, we asked them to evaluate the degree to which they believed their knowledge affected their decisions. In this section, we aim to compare their self-reported influence with their actual choices by conducting chi-square tests and analyzing cross-tabulations. The results, presented in Fig.20 and Fig.21, indicate that there is no statistically significant relationship between participants' perceived influence and their actual query choices. In other words, the statistical analysis shows no clear correlation between participants' self-assessed influence and their actual preferences for linear or directional top-k queries, highlighting a mismatch between perceived and actual impact.
\begin{figure}
    \centering
    \begin{subfigure}[b]{0.5\textwidth}
        \includegraphics[width=\textwidth]{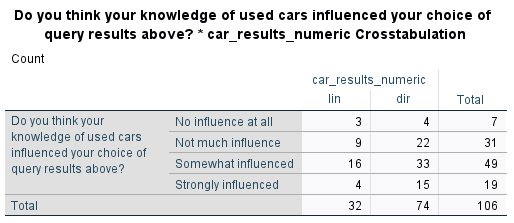}
        \caption{Crosstabulation Result}
        \label{fig:Crosstabulation Result}
    \end{subfigure}
    \hfill
    \begin{subfigure}[b]{0.5\textwidth}
        \includegraphics[width=\textwidth]{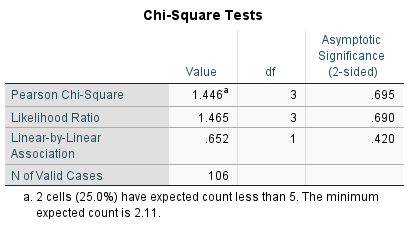}
        \caption{Chi-Square result}
        \label{fig:Chi-Square result}
    \end{subfigure}
    \caption{Chi-Square test for self-report influence on user preference in used cars topic}
    \label{fig:Chi-Square test for self-report influence on user preference in used cars topic}
\end{figure}

\begin{figure}
    \centering
    \begin{subfigure}[b]{0.5\textwidth}
        \includegraphics[width=\textwidth]{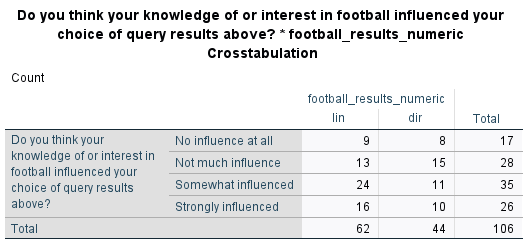}
        \caption{Crosstabulation Result}
        \label{fig:Crosstabulation Result}
    \end{subfigure}
    \hfill
    \begin{subfigure}[b]{0.5\textwidth}
        \includegraphics[width=\textwidth]{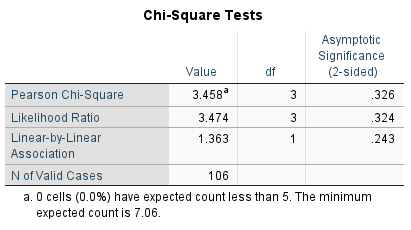}
        \caption{Chi-Square result}
        \label{fig:Chi-Square result}
    \end{subfigure}
    \caption{Chi-Square test for self-report influence on user preference in football player topic}
    \label{fig:Chi-Square test for self-report influence on user preference in football player topic}
\end{figure}

\subsection{Summary of findings}
Through the analysis, several key findings have emerged. Within the two topics examined, a significant preference for the improved directional top-k query was identified in the "used cars" topic, partially supporting Hypothesis 1. Conversely, in the "football players" topic, while 58\% of users preferred linear queries, this preference was not statistically significant, suggesting no clear bias toward either query type. Additionally, the statistically significant difference in user preferences between the two topics confirms that the datasets should not be analyzed together, further supporting Hypothesis 2 that topic type influences user preferences. Specifically, in the "used cars" topic, users were 3.259 times more likely to choose directional queries than in the "football players" topic. 

Regarding knowledge and experience, these variables had no significant impact on preferences in the "used cars" topic but did exhibit a significant effect in the "football players" topic. Participants with higher levels of football knowledge (advanced and expert) demonstrated a stronger preference for linear queries, whereas those with lower levels of knowledge (beginner and intermediate) displayed more balanced preferences between the two query types. As for gender and age, neither variable significantly influenced preferences in the "used cars" topic. However, in the "football players" topic, age showed marginal significance, with a potential linear relationship between age and query preference. Lastly, participants' self-reported influence of their knowledge on their choices did not align with their actual preferences. Statistical analysis revealed no significant correlation between perceived and actual impacts, highlighting a disconnect between subjective perceptions and actual behavior.

\subsection{Interpretation of Key Findings}

In the previous section, we concluded that user preferences for query methods are influenced by the topic, and the two dataset types selected for this study significantly impact user preferences. For lifestyle-related topics closely tied to daily life, such as those aimed at e-commerce and online shopping, users exhibit a clear preference for directional queries. This approach retrieves more balanced results, helping users identify products that better meet their expectations, such as those with greater cost-effectiveness. Conversely, in specialized areas such as hobbies or interests, user preferences for query rankings appeared more random, indicating that their choices are driven more by subjective factors.

 we interviewed two participants about their preferences for football players topic. One advanced-level football enthusiast explained his/her preference for linear top-k queries, stating that the players retrieved in these queries were all familiar to them and largely aligned with his/her perception of top players. For him/her, the top 10 players should reflect this specific ranking. In contrast, a beginner-level respondent with no prior football experience expressed a belief that football, being a highly collaborative sport, does not require balance across abilities for players to rank at the top. Instead, exceptional performance in a specific skill could qualify a player as a top performer, as reflected in linear query results. These observations suggest that users engaging with specialized topics might favor results that focus on specific attributes rather than balanced outcomes.

 The analysis of knowledge and experience demonstrated that, for lifestyle-related topics, factors such as gender, age, and background knowledge had no significant impact. Users’ primary goal in these contexts is finding products that fulfill their needs, emphasizing the utility of directional queries in such scenarios. However, for football-related topics, age and knowledge were shown to have some influence, with age displaying a marginally significant linear relationship with query preferences. Older users seemed to leverage their life experiences to prioritize certain attributes, while users with advanced knowledge were more likely to appreciate rankings produced by linear queries, which often aligned with their expertise and expectations.

 These findings further suggest that for professional or interest-driven topics, user preferences are shaped by subjective goals rather than a desire for balanced results. Highly knowledgeable users may use queries to confirm their preconceived notions, while less knowledgeable users depend on the results for guidance. This discrepancy could be attributed to differences in their ability to interpret data, leading to diverse preferences for query ranking methods.

\section{Conclusion and Future work}

\subsection{Limitation}
This research stands out by employing real user data, offering an authentic insight into user preferences for query methods compared to previous studies. However, it also has certain limitations. The small sample size (N=106) represents a major constraint, potentially limiting the statistical power of the analyses. For example, while age appeared to have a marginally significant linear effect on preferences within the football topic, the limited sample size restricted the reliability of this conclusion. Additionally, the sample predominantly included individuals under 35 years of age, with limited representation from older demographics, making it challenging to generalize findings to older populations.

In terms of survey design, simplicity was prioritized to help participants better understand the experiment's objectives and maintain their focus. As such, only two attributes were selected as the basis for query rankings. However, real-world query rankings often involve numerous attributes and different weights, users' evaluation criteria can vary significantly. Furthermore, the experiment always uses the same weights for the selected attributes, which may not fully reflect situations where attributes have different degrees of importance to users. These simplifications may limit the study's ability to fully replicate real-world decision-making contexts and capture nuanced user preferences.

Additionally, the selection of topics—used cars for e-commerce and football players for specialized interests—may not comprehensively represent these categories or resonate equally with all participants. This study only compared two specific topics to analyze differences in user preferences across topic types. While the findings reveal significant differences in user preferences for these particular topics, they cannot be generalized to all possible topic categories or contexts. Finally, the self-reported nature of participants' knowledge levels and perceived influence on their choices introduces a degree of subjectivity, which may have influenced the study's results.

\subsection{Future work}

Future research on user preferences for top-k queries should aim to include larger and more diverse topics and samples, with a particular focus on increasing the representation of individuals aged 35 and above. This expansion will enhance the applicability of the findings and provide a more comprehensive understanding of query preferences across different age groups. Furthermore, incorporating more complex and realistic query tasks, such as introducing multiple attributes into the ranking process, could provide valuable insights into whether task complexity affects user preferences. Additionally, experimenting with unbalanced weight distributions might offer a deeper understanding of how varying attribute importance influences user preference. Moreover, future studies could adopt a mixed-methods approach by integrating quantitative research with qualitative methods, such as conducting in-depth interviews with participants to explore their preferences and decision-making processes in greater detail.

In addition, future research could further explore the application of directional queries in lifestyle-related topics, such as the development prospects of e-commerce and users' preferences. This direction could provide valuable insights into how directional query methods can be tailored to meet the dynamic needs of users in various lifestyle scenarios, enriching the practical implications of such queries.

\subsection{Conclusion}

This paper provides a comparative analysis of user preferences for Linear Top-k Query and Directional Top-k Query methods, focusing on two distinct topics: used cars in the e-commerce domain and football players in the personal interest domain. To capture real user perspectives, the study employed a structured questionnaire, allowing participants to evaluate query results and share their preferences in a realistic and user-centric method. The findings reveal a significant preference for Directional Queries in lifestyle-related topics like used cars, where balanced results align with user needs. In contrast, preferences in specialized topics like football players were more evenly distributed. 

For the used cars topic, background knowledge, experience, age, and gender had no significant impact, highlighting the broad appeal and practicality of directional queries for everyday applications. For the football players topic, preferences were more evenly distributed, with knowledgeable users favoring Linear Queries that emphasize extreme attribute values, while less knowledgeable users showed more balanced preferences. Demographic factors, such as age, displayed a marginal influence, suggesting that user expertise and individual goals play a larger role in shaping preferences in specialized domains.

In conclusion, this research demonstrates the potential for Directional Queries to enhance user satisfaction in lifestyle-related applications, while also illustrating the subjective nature of preferences in specialized domains. Despite these contributions, the study has limitations, including a small sample size, underrepresentation of individuals aged 35 and above, and a simplified survey design that focused on two attributes with equal weights and specific topics. These factors may restrict the applicability of the findings and the replication of real-world decision-making contexts. Future research should address these limitations by incorporating larger and more diverse samples, exploring more complex, multi-attribute query tasks, and adopting mixed-method approaches such as in-depth interviews to provide deeper insights into user preferences and decision-making processes.

\bibliographystyle{unsrt}  
\bibliography{references}  

\end{document}